\documentclass[twocolumn]{aastex631}

\usepackage[shortlabels]{enumitem}
\usepackage{nicefrac}

\begin{document}

\title{Massive Star Cluster Formation with Binaries. I. Evolution of Binary Populations}

\correspondingauthor{Claude Cournoyer-Cloutier}
\email{cournoyc@mcmaster.ca}

\author[0000-0002-6116-1014]{Claude Cournoyer-Cloutier}
\affiliation{Department of Physics and Astronomy, McMaster University, 1280 Main Street West, Hamilton, ON, L8S 4M1, Canada}

\author[0000-0003-3551-5090]{Alison Sills}
\affiliation{Department of Physics and Astronomy, McMaster University, 1280 Main Street West, Hamilton, ON, L8S 4M1, Canada}

\author[0000-0001-8762-5772]{William E. Harris}
\affiliation{Department of Physics and Astronomy, McMaster University, 1280 Main Street West, Hamilton, ON, L8S 4M1, Canada}

\author[0000-0001-5972-137X]{Brooke Polak}
\affiliation{Zentrum f\"{u}r Astronomie, Institut f\"{u}r Theoretische Astrophysik, Universit\"{a}t Heidelberg, Albert-Ueberle-Str. 2, 69120 Heidelberg, Germany}
\affiliation{Department of Astrophysics, American Museum of Natural History, 200 Central Park West, New York, NY 10024-5102, USA}

\author[0000-0003-3688-5798]{Steven Rieder}
\affiliation{Anton Pannekoek Institute for Astronomy, University of Amsterdam, Science Park 904, 1098 XH Amsterdam, The Netherlands}
\affiliation{Institute of Astronomy, Department of Phyiscs and Astronomy, KU Leuven,
Celestijnenlaan 200D bus 2401, 3001 Leuven, Belgium}

\author[0000-0003-3479-4606]{Eric P. Andersson}
\affiliation{Department of Astrophysics, American Museum of Natural History, 200 Central Park West, New York, NY 10024-5102, USA}

\author[0000-0002-6593-3800]{Sabrina M. Appel}
\thanks{NSF Astronomy and Astrophysics Postdoctoral Fellow}
\affiliation{Department of Astrophysics, American Museum of Natural History, 200 Central Park West, New York, NY 10024-5102, USA}
\affiliation{Department of Physics and Astronomy, Rutgers University, 136 Frelinghuysen Road,
Piscataway, NJ 08854-8019, USA}

\author[0000-0003-0064-4060]{Mordecai-Mark Mac Low}
\affiliation{Department of Astrophysics, American Museum of Natural History, 200 Central Park West, New York, NY 10024-5102, USA}

\author[0000-0001-9104-9675]{Stephen McMillan}
\affiliation{Department of Physics, Drexel University, 3141 Chestnut Street, Philadelphia, PA 19104, USA}

\author[0000-0001-5839-0302]{Simon Portegies Zwart}
\affiliation{Sterrewacht Leiden, Leiden University, Einsteinweg 55, 2333CC Leiden, The Netherlands}



\begin{abstract}
We study the evolution of populations of binary stars within massive cluster-forming regions. We simulate the formation of young massive star clusters within giant molecular clouds with masses ranging from 2 x 10$^{4}$ to 3.2 x 10$^{5}$ M$_{\odot}$. We use \textsc{Torch}, which couples stellar dynamics, magnetohydrodynamics, star and binary formation, stellar evolution, and stellar feedback through the \textsc{Amuse} framework. We find that the binary fraction decreases during cluster formation at all molecular cloud masses. The binaries' orbital properties also change, with stronger and quicker changes in denser, more massive clouds. Most of the changes we see can be attributed to the disruption of binaries wider than 100 au, although the close binary fraction also decreases in the densest cluster-forming region. The binary fraction for O stars remains above 90\%, but exchanges and dynamical hardening are ubiquitous, indicating that O stars undergo frequent few-body interactions early during the cluster formation process. Changes to the populations of binaries are a by-product of hierarchical cluster assembly: most changes to the binary population take place when the star formation rate is high and there are frequent mergers between sub-clusters in the cluster-forming region. A universal primordial binary distribution based on observed inner companions in the Galactic field is consistent with the binary populations of young clusters with resolved stellar populations, and the scatter between clusters of similar masses could be explained by differences in their formation history. 
\end{abstract}



\section{Introduction} \label{sec:intro}

Young massive clusters (YMCs) are ubiquitous in nearby star-forming galaxies, with 
populations of massive clusters 
embedded in their natal gas observed in nearby starbursts~\citep[e.g.,][]{He2022, Sun2024}. 
Observations have also found massive and compact star clusters at high redshift (e.g., $z$$\sim$6, \citeauthor{Vanzella2023}~\citeyear{Vanzella2023}; $z$$\sim$10.2, \citeauthor{Adamo2024}~\citeyear{Adamo2024}), 
and 
globular clusters (GCs) at $z$=1.38~\citep[e.g.,][]{Claeyssens2023, Adamo2023} and at $z$$\sim$0.3-0.4~\citep[e.g.,][]{Faist2022, Lee2022, Harris2023}. 
These observations serve as growing evidence that GC formation is the extension at high masses of YMC formation in the local Universe. This is also supported by simulations, which have shown 
that star clusters of masses $\gtrsim$ 10$^6$ M$_{\odot}$ form naturally within massive giant molecular clouds~\citep[e.g.,][]{Howard2018, Polak2024, Reina-Campos2024}. 
%
Detailed analysis of the stellar populations of YMCs and the giant molecular clouds (GMCs) within which they form are limited to the Milky Way and the LMC/SMC, for which the cluster mass function does not reach the 
high masses observed in starbursts~\citep{PortegiesZwart2010}. This is a challenge for understanding the spatial and kinematic properties of the stars, including the binary fraction,
which are 
largely unknown for the most massive YMCs.

Massive stars, which regulate star formation within their host cluster, have binary fractions approaching unity~\citep{Moe2017} 
and two-thirds of all massive stars exchange material with a companion over their lifetimes~\citep{Sana2012}. This changes the time distribution of supernovae~\citep{Zapartas2017}, increases the amount of ultraviolet radiation~\citep{Goetberg2018} and increases the amount of pre-supernova ejecta from a stellar population by a factor of $\sim$6~\citep{Nguyen2024}. Binaries 
can produce runaway stars either through the ejection of a newly-unbound companion after a supernova~\citep{Blaauw1961}, or through few-body interactions~\citep{Poveda1967}. Such runaway stars are known to be ubiquitous around young clusters~\citep[e.g.,][]{Kalari2019, Stoop2023, Stoop2024}, and in turn influence the distribution of feedback within cluster-forming regions which can affect the long-term evolution of galaxies~\citep{Andersson2020, Andersson2023, Steinwandel2023}.

Clustered environments also influence the binaries found within them: high stellar densities promote few-body interactions that can result in the formation, modification, or disruption of binaries. Observationally, surveys of binaries in clustered environments 
suggest a dependence of the binary fraction and orbital properties of binaries on cluster density and/or cluster mass. The frequency and orbital properties of close ($<$ 10 au) companions to low- and solar-mass stars appears to be set very early in the stars' evolution~\citep[e.g.,][]{Kounkel2019} and to persist both in young open clusters~\citep[e.g.,][]{Deacon2020} and in the field, suggesting that the field properties observed for those systems resemble 
the primordial ones. On the other hand, the wide binary fraction depends on environment
: low-density star forming regions show an excess of wide companions~\citep[e.g., Taurus,][]{Kraus2011, Joncour2017} while high-density star-forming regions show fewer wide companions
~\citep[e.g., Orion,][]{Duchene2018, Jerabkova2019}. 
It was suggested that the binary fraction observed in the field arises 
from a combination of binary fractions inherited from stars formed in embedded clusters of varying densities, which have since dissolved~\citep[see][for a discussion]{Offner2023}. Observations of older clusters do not fully support this picture. For open clusters, surveys have found evidence of binary fraction either increasing~\citep[e.g.,][]{Niu2020} or decreasing~\citep{Deacon2020} with stellar density, depending on which clusters were observed and what separation range was probed. The older, denser, more massive GCs consistently show low binary fractions, which are anti-correlated with cluster mass but do not, however, show any trends with cluster density~\citep[e.g.,][]{Milone2012}. Taken together, these results indicate that there is more at play than just the present-day densities for setting the binary fraction and binary properties.

%
We need to understand how a binary population evolves during the formation of clusters at a range of masses and densities in order to fully understand cluster formation and long-term evolution. 
Numerical simulations provide us with detailed 
spatial and kinematic 
information about each star within a YMC, as a function of time. Modelling the stars alongside gas, which is affected by feedback, is crucial: populations of binaries are modified during cluster assembly~\citep{Cournoyer-Cloutier2021}, and these changes are driven by subcluster mergers~\citep{Cournoyer-Cloutier2024}, which are in turn dependent on the GMC-scale gas environment~\citep{Lahen2020, Guszejnov2022, Karam2024}. 

We present a suite of simulations with cloud masses ranging from 2 x 10$^4$ to 3.2 x 10$^5$ M$_{\odot}$ to test the dependence of binary properties on environment during star cluster formation. Those simulations include primordial binaries, star formation, and stellar feedback along with collisional stellar dynamics and magnetohydrodynamics. We describe our simulation methods in Section~\ref{sec:methods} and our suite of simulations in Section~\ref{sec:simulations}. We present our results in Section~\ref{sec:results} and discuss their implications for globular cluster formation in Section~\ref{sec:discussion}. We conclude in Section~\ref{sec:conclusions}. 

\section{Methods} \label{sec:methods}
We perform our simulations with \textsc{Torch}\footnote{\url{https://bitbucket.org/torch-sf/torch/src/binaries-v2.0/}}~\citep{Wall2019, Wall2020}, which couples magnetohydrodynamics (MHD) to star formation, stellar dynamics, stellar evolution, and stellar feedback through the \textsc{Amuse} framework~\citep{PortegiesZwart2009, Pelupessy2013, PortegiesZwart2013, PortegiesZwart2019}. The coupling between the different codes is presented in~\citet{Wall2019}.
The physics in our simulations are described in more detail in the following sections. Parameters for star formation, feedback, N-body dynamics and binary formation, which are shared between the simulations, are summarized in Table~\ref{tab:numerics}. 

\begin{table}[tb!]
    \begin{center}
    \begin{tabular}{lll}
     & Definition & Value \\
    \hline
    $\Delta x_{\mathrm{min}}$ & Minimum cell size & 0.137 pc \\
    $r_{\mathrm{sink}}$ & Sink particle accretion radius & 0.342 pc \\
    $\rho_{\mathrm{sink}}$ & Gas density for sink formation & 93.5 M$_{\odot}$ pc$^{-3}$\\
    $M_{\mathrm{sink}}$ & Minimum sink mass at formation & 3.74 M$_{\odot}$\\
    $T_{\mathrm{wind}}$ & Wind target temperature & 3 x 10$^5$ K\\
    \hline
    $r_{\mathrm{out}}$ & Changeover radius for \textsc{fdps} & Section~\ref{subsec:petar} \\
    $r_{\mathrm{in}}$ & Changeover radius for N-body & 0.1 $r_{\mathrm{out}}$ \\
    $r_{\mathrm{bin}}$ & Changeover radius for \textsc{sdar} & 100 au \\
    \hline
    $M_{\mathrm{min}}$ & Minimum stellar mass & 0.4 M$_{\odot}$ \\
    $M_{\mathrm{max}}$ & Maximum stellar mass & 150 M$_{\odot}$ \\
    $M_{\mathrm{FB}}$ & Minimum mass for feedback & 13 M$_{\odot}$ \\
    $\mathcal{F}_{\mathrm{bin}}$ & Binary fraction & Section~\ref{subsec:primordial binaries} \\
    \hline
    \end{tabular}
    \end{center}
    \caption{Parameters for the simulations. $r_{\mathrm{out}}$ depends on the simulation timestep and is described in Section~\ref{subsec:petar}. $\mathcal{F}_{\mathrm{bin}}$ depends on the stellar mass and is described in Section~\ref{subsec:primordial binaries}.}
    \label{tab:numerics}
\end{table}

\subsection{Magnetohydrodynamics}\label{subsec:flash}
We use \textsc{Flash}~\citep{Fryxell2000, Dubey2014} with a Harten-Lax-van Leer Riemann solver~\citep{Miyoshi2005} with third-order piecewise parabolic method reconstruction~\citep{Colella1984}. We use a multigrid solver~\citep{Ricker2008} for the gas self-gravity. Gravity between the gas and stars is treated with a leapfrog scheme based on \textsc{Bridge}~\citep{Fujii2007}. 
We refine our adaptive grid such that the Jeans length is resolved by at least 12 resolution elements, to ensure that it can be magnetically supported against collapse on scales below the resolution (\citeauthor{Heitsch2001}~\citeyear{Heitsch2001}, see also discussion in~\citeauthor{Federrath2010}~\citeyear{Federrath2010}). To improve numerical stability in regions with large temperature or pressure gradients, such as HII regions, we also refine where the second derivative of the temperature or pressure is of the order of the sum of its gradients~\citep[see][]{Lohner1987, MacNeice2000}. 
Sink particles are used to model sub-grid star and binary formation. Sinks form in regions of high gas density and converging flows which satisfy the boundedness and gravitational instability criteria outlined in~\citet{Federrath2010}. The sink accretion radius is set to 2.5$\Delta x$ at the highest refinement level, and sinks can only form in regions that are refined to the highest refinement level. We give more details on the formation of stars from sinks in Section~\ref{subsec:feedback}. 

\subsection{Stellar dynamics}\label{subsec:petar}
We handle stellar dynamics, including hard binaries and close encounters, with the N-body code \textsc{PeTar}~\citep[][see \citealt{Polak2024} for the implementation in \textsc{Torch}]{Wang2020b}. \textsc{PeTar} relies on a combination of three different N-body algorithms: long-range interactions are calculated with a Barnes-Hut tree~\citep[][as implemented in \textsc{fdps} by~\citealt{Iwasawa2016}]{Barnes1986}, short-range interactions are calculated with a fourth-order Hermite integrator~\citep{Makino1992}, and binary systems and few-body encounters are calculated with the slow-down algorithmic regularization method~\citep[\textsc{sdar}, ][]{Wang2020a}. 

The changeover radii between these three integration regimes are set by the user. The default values calculated by \textsc{PeTar} are optimized for a spherical stellar system with a fixed number of stars. However, our cluster-forming regions continuously form new stars, and are not well-described by a single spherical cluster at early times. They also span a range of densities. As we are interested in the evolution of binaries within these fast-changing stellar systems, we adopt larger changeover radii than previous work using \textsc{PeTar}: this is less efficient than optimizing \textsc{PeTar} for binaries expected to survive the cluster's long-term evolution, but ensures that we fully capture the complexity of stellar dynamics within the cluster-forming region. Binary systems on scales smaller than $r_{\mathrm{bin}}$ are treated by \textsc{sdar}, ensuring that they do not affect the global timestep of the simulation. Forces between stars closer than $r_{\mathrm{in}}$ are calculated using the fourth-order Hermite, forces between stars more distant than $r_{\mathrm{out}}$ are calculated using the Barnes-Hut tree, and forces between stars with separations between $r_{\mathrm{in}}$ and $r_{\mathrm{out}}$ are treated using a combination of the Hermite and tree codes. 

The changeover radius for \textsc{sdar} $r_{\mathrm{bin}}$ is kept fixed in our simulations.  We adopt a value of 100 au as this corresponds to an orbital period of 
58 years for a binary with two stars of mass 150 M$_{\odot}$ (the largest stellar mass in our simulations). As a typical timestep for our simulations is between 15.625 years and 62.5 years, this ensures that the orbits of massive binaries (with large orbital velocities) are well-resolved in our simulations. 100 au is also commonly used as the lower semi-major axis limit for wide binaries in observational surveys~\citep[see][and references therein]{Offner2023}.

The changeover radii $r_{\mathrm{in}}$ and $r_{\mathrm{out}}$ have a strong impact on the performance of the code but must be chosen carefully in conjunction with the simulation timestep to ensure that the orbits of binaries wider than 100 au but still strongly bound are well-resolved.  \textsc{Torch} uses a single timestep for MHD and stellar dynamics; as we model star-forming regions with shock fronts and high sound speeds, the timestep calculated by \textsc{Flash} decreases drastically after the formation of the first massive star. On the other hand, the \textsc{PeTar} timestep must always remain a power of 2 of the initial \textsc{PeTar} timestep, which we set to 1000 years. The only allowed timesteps for the simulation are therefore 
\begin{equation}
    \Delta t = \frac{1000}{2^n} \,\, \mathrm{years}
\end{equation}
where $n$ is a positive integer. After the formation of the first star, we set the maximum timestep and $r_{\mathrm{out}}$ together, such that the orbit of a circular binary with a semi-major axis of $r_{\mathrm{out}}$ is resolved by 10 timesteps if the two stars have a mass of 10 M$_{\odot}$. For our simulations, the shortest minimum timestep used is 7.8125 years and the longest minimum timestep used is 125 years, which correspond respectively to $r_{\mathrm{out}}$ = 0.00112~pc and $r_{\mathrm{out}}$ = 0.00709~pc. The inner changeover radius $r_{\mathrm{in}}$ is set to 0.1 $r_{\mathrm{out}}$ at all times, following the standard approach in \textsc{PeTar}. The computational time is insensitive to our choice of \textsc{PeTar} timestep, as the computational time per time step is dominated by \textsc{FLASH} for all timesteps used in the paper.

\subsection{IMF sampling}\label{subsec:IMF}
Star formation takes place via sink particles from which individual stars are spawned. This process is described in details in~\citet{Wall2019} and the modifications to this method to allow for primordial binary formation are described in~\citet{Cournoyer-Cloutier2021}. 
As each sink is formed, we sample a~\citet{Kroupa2002} initial mass function (IMF) and apply a primordial binary prescription, described in Section~\ref{subsec:primordial binaries}, to generate a list of stars to be formed. 
We use a lower limit of 0.4 M$_{\odot}$ and an upper limit of 150 M$_{\odot}$ to sample the IMF. The lower limit reduces the number of stars by a factor of 2 compared to sampling down to 0.08 M$_{\odot}$, reducing the load on the N-body integrator while retaining 90\% of the stellar mass in stars of the mass predicted by the IMF. Each sampled star corresponds to a star particle, to ensure that the shape of the IMF is preserved. Although low-mass binaries are important to the long-term evolution of star clusters once massive stars have evolved, they are not the leading source of binding energy in young cluster-forming regions, which host several massive stars for which the close binary fraction approaches unity. 
We inject radiative and wind feedback from all stars more massive than 13~M$_{\odot}$.

\subsection{Primordial binaries}\label{subsec:primordial binaries}

We use the module for primordial binaries in \textsc{Torch} that was first presented in \citet{Cournoyer-Cloutier2021}, in which we implement an updated sampling binary algorithm. We use a mass-dependent binary fraction based on observations made by~\citet{Winters2019} and corrected by~\citet[][]{Offner2023} for stars below 0.6~M$_{\odot}$, and on observations compiled by~\citet[][]{Moe2017} for stars above 0.8~M$_{\odot}$; the only difference between those fractions and those used in~\citet{Cournoyer-Cloutier2021} is the correction by~\citet{Offner2023} of the binary fraction for M$\leq$ 0.6~M$_{\odot}$. The sampling technique for the orbital period, companion mass, and eccentricity also remains the same as in~\citet{Cournoyer-Cloutier2021}.

\begin{table}[tb!]
    \begin{center}
    \begin{tabular}{ccccc}
    Mass range \,\, & \,\, $\mathcal{F}_{\mathrm{bin}}$ \,\, & \,\, $\mathcal{F}_{\mathrm{close}}$ \,\, & \,\, a$_{\mathrm{median}}$ \,\, & \,\, q$_{\mathrm{median}}$ \\
    \hline
    0.4-0.8 M$_{\odot}$ & 0.30 & 0.07 & 44.6 au & 0.90 \\
    0.8-1.6 M$_{\odot}$ & 0.40 & 0.15 & 201 au & 0.67\\
    1.6-5 M$_{\odot}$ & 0.59 & 0.37 & 21.6 au & 0.41\\
    5-9 M$_{\odot}$ & 0.76 & 0.63 & 9.92 au & 0.34\\
    9-16 M$_{\odot}$ & 0.84 & 0.80 & 7.08 au & 0.35\\
    $\geqslant$ 16 M$_{\odot}$ & 0.94 & 0.94 & 6.72 au & 0.37\\
    \hline
    \end{tabular}
    \end{center}
    \caption{Binary fraction, close binary fraction, median semi-major axis and median mass ratio, for each mass range. Those values are calculated for a fully sampled distribution of binaries.}
    \label{tab:frac_ICs}
\end{table}

The key change in the updated sampling algorithm is in the distribution of orbital periods.  The sampling algorithm used in~\citet{Cournoyer-Cloutier2021} sampled \textit{any} companion from a distribution of observed properties, while the new algorithm presented here is designed to sample the \textit{inner} companion of a hierarchical multiple stellar system. The triple fraction ranges from 10\% for solar-mass stars to 73\% for O-type stars~\citep{Moe2017}, which implies that the distribution of \textit{all companions} to O stars is quite different from the distribution of \textit{inner companions} to O stars. The updated algorithm therefore accounts for the dynamical formation of hierarchical triples and higher-order multiples while preserving the observed close binary fraction. For stars with masses above 0.8 M$_{\odot}$, we impose that a fraction $\mathcal{F}_{\mathrm{close}}$ of all stars in each mass range must have a companion with an orbital period shorter than 5000 days. For a binary of total mass 100 M$_{\odot}$, this corresponds to a semi-major axis of roughly 27 au while it corresponds to a semi-major axis of about 10 au for a binary with total mass 5 M$_{\odot}$. For stars with masses below 0.8 M$_{\odot}$, we use the lognormal distribution of semi-major axes for inner companions from~\citet{Winters2019}, as reported and corrected in~\citet{Offner2023}, which corresponds to a mean semi-major axis of 14 au for close binaries. A fraction $\mathcal{F}_{\mathrm{bin}} - \mathcal{F}_{\mathrm{close}}$ of stars in each mass bin will have a companion with an orbital period longer than 5000 days.
The binary fraction, close binary fraction, median semi-major axis, and median mass ratio for each mass bin are reported in Table~\ref{tab:frac_ICs}.

The sampling algorithm only forms systems that do not fill their Roche lobe while on the zero-age main sequence. This is done using an upper limit on the eccentricity as a function of period based on the semi-analytic formula from~\citet{Moe2017}, as described in~\citet{Cournoyer-Cloutier2021},
\begin{equation}
    e_{\mathrm{max}} (P) = \Big{(}\frac{P}{2 \, \mathrm{days}}\Big{)}^{\nicefrac{-2}{3}}
\end{equation}
where $P$ is the orbital period in days. Stars are allowed to merge during the simulations, but other binary evolution effects -- such as stable or unstable mass transfer -- are not taken into account. 



\subsection{Feedback}\label{subsec:feedback}
Radiation from stars more massive than 13~M$_{\odot}$ is followed using the ray-tracing scheme \textsc{Fervent}~\citep{Baczynski2015}, which follows radiation pressure in the far-ultraviolet band (FUV, 5.6-13.6 eV) and ionizing radiation (above 13.6 eV) from individual massive stars. Momentum-driven winds are also injected on the grid by massive stars. The implementation of both forms of feedback is described in~\citet{Wall2020}. \textsc{Torch} also includes a scheme for core-collapse supernovae, implemented within \textsc{Flash}; our simulations however stop before any supernovae takes place. 
We mass-load our winds by increasing the wind mass loss rate $\dot{M}$ while keeping the wind luminosity $L_w$ fixed, therefore lowering the wind velocity $v_w$, following
\begin{equation}
    L_w = \frac{1}{2} \dot{M} v_w^2 .
\end{equation}
The wind velocity is reduced to reach a target post-shock wind temperature
\begin{equation}
    T_w = 1.38 \, \mathrm{x} \, 10^7 \, \mathrm{K} \, \Bigg{(} \frac{v_w}{10^3 \, \mathrm{km} \, \mathrm{s}^{-1}}\Bigg{)}^2
\end{equation}
following~\citet{Wall2020}. We mass-load the winds to a target temperature $T_{w}$ = 3 x 10$^5$ K, which was shown to be a reasonable choice in~\citet{Polak2024} and allows for a longer timestep. Mass loading makes the wind bubbles momentum-driven rather than energy-driven. Momentum-driven winds are naturally produced by high-resolution hydrodynamics wind simulations~\citep[see e.g.,][]{Lancaster2021}: mass-loading our winds allows us to reproduce the effects of the winds on cluster-scale at the resolution of our simulations.

\section{Overview of simulations}\label{sec:simulations}

\subsection{Initial conditions}\label{subsec:ICs}

\begin{figure*}[tb!]
    \centering
    \includegraphics[width=\linewidth, clip=True, trim=3cm 0cm 0cm 11cm]{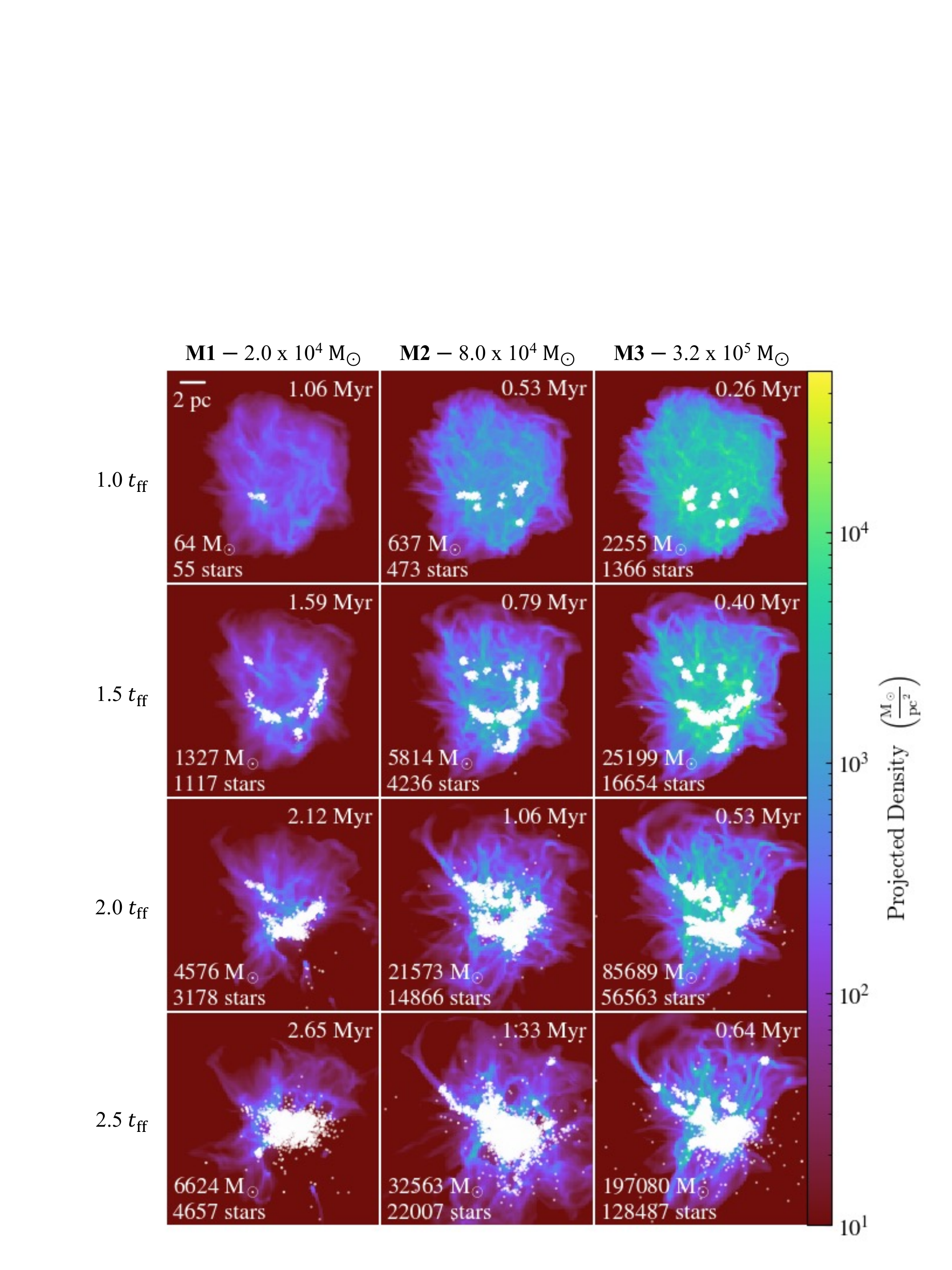}
    \caption{Gas column density, with stars overplotted as small white circles. The columns correspond to the different simulations while the rows show the distribution of stars and gas at approximately 1.0, 1.5, 2.0 and 2.5 free-fall times of the initial GMC.}
    \label{fig:Gas projectections}
\end{figure*}

We run a suite of isolated cloud simulations with initial gas masses of 2 x 10$^4$, 8 x 10$^4$, and 3.2 x 10$^5$ M$_{\odot}$ and an initial cloud radius $R=7$ pc. Each model is run for at least 2.5 free-fall times (calculated for the initial cloud), with the lowest mass model run until gas expulsion. As GMC masses correlate with GMC surface densities but show little to no correlation with GMC radius or virial parameter for extragalactic GMCs~\citep{Sun2022}, we vary the initial GMC mass but keep the radius fixed, which changes the surface density and the density between simulations. The surface densities for M1, M2 and M3 are chosen respectively to mimic GMC conditions typical of the disk of the Milky Way~\citep[e.g.,][]{Roman-Duval2010, Chen2020}, the Central Molecular Zone or centers of barred galaxies~\citep[e.g.,][]{Sun2020}, and starburst galaxies~\citep[e.g.,][]{Sun2018}.
All simulation domains have a box side of $L = 2.5R = 17.5$~pc. All models are initialized with an initial virial parameter $\alpha = 2T/|U| = 0.5$, a Kolmogorov turbulent velocity spectrum with the same random seed, and a uniform magnetic field $B_z = 3 \, \mu G$. The initial GMCs are isolated and the simulations do not include any external tidal field, which would not affect the clusters' evolution on the short ($<$ 5 Myr) timescales considered~\citep[see][]{Miholics2017}. The initial conditions are summarized in Table~\ref{tab:ICs}. 
The gas column density overlaid with the stellar distribution is presented in Figure~\ref{fig:Gas projectections} for all simulations, shown at the same fractions of their respective initial free-fall times. 

%

\begin{table*}[tb!]
    \begin{center}
    \begin{tabular}{lccccccccccc}
    Cloud & M$_{\mathrm{gas}}$ [M$_{\odot}$]  & $R$ [pc] & $\Sigma$ [M$_{\odot}$ pc$^{-2}$] & $t_{\mathrm{ff}}$ [Myr] & SFE [2.5 $t_{\mathrm{ff}}$] & $F_{\mathrm{bound}}$ [2.5 $t_{\mathrm{ff}}$] & SFE & $M_{\mathrm{formed}}$ [M$_{\odot}$] & t$_{\mathrm{SFR}}$ \\
    \hline
    M1 & 2 x 10$^4$ & 7 & 130 & 1.06 & 0.33 & 0.98 & 0.39 & 7.8 x 10$^3$ &1.56\\
    M2 & 8 x 10$^4$ & 7 & 520 & 0.530 & 0.40 & 0.99 & $\geqslant$ 0.43 & 3.44 x 10$^4$ &  1.64 \\
    M3 & 3.2 x 10$^5$ & 7 & 2080 & 0.265 & 0.61 & 0.99 & $\geqslant$ 0.61 & 1.95 x 10$^5$ & $\gtrsim$ 2.46\\
    \hline
    \end{tabular}
    \end{center}
    \caption{Initial conditions and star formation metrics for the simulations. Columns: Cloud label, initial mass, initial radius, initial surface density, and initial free-fall time of the gas cloud; star formation efficiency after 2.5 $t_{\mathrm{ff}}$, bound mass fraction after 2.5 $t_{\mathrm{ff}}$, total star formation efficiency, formed stellar mass, and time at which the star formation rate peaks (in units of free-fall times).}
    \label{tab:ICs}
\end{table*}

\subsection{Star formation} 

\begin{figure}[tb!]
    \centering
    \includegraphics[width=\linewidth, clip=True, trim=0cm 0cm 0cm 0cm]{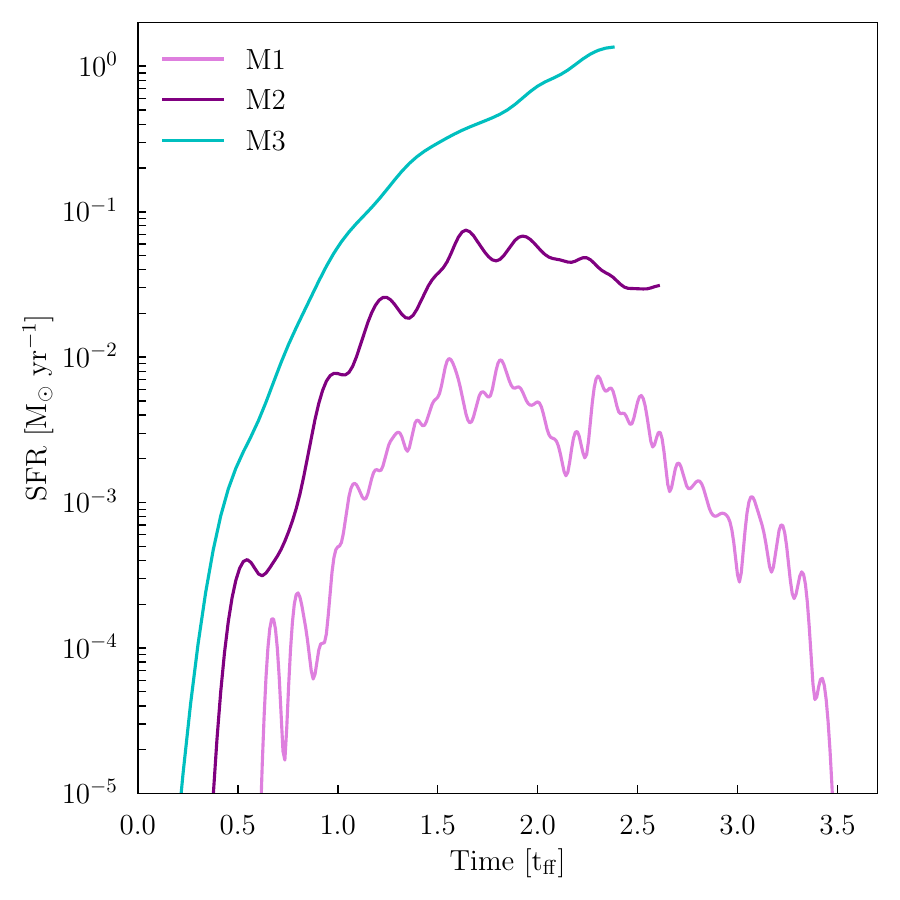}
    \caption{Star formation rate as a function of time in units of initial GMC free-fall time. Note that all lines are slightly truncated at the end due to the Gaussian smoothing.}
    \label{fig:SFR}
\end{figure}

We summarize the key star formation metrics in Table~\ref{tab:ICs}, and plot the star formation rate (SFR) as a function of time in Figure~\ref{fig:SFR}. As expected, the star formation rate increases with initial gas mass, but in excess of the increase in mass -- in other words, the increased total gas mass results in more stars formed, but the increased initial surface density also increases both the SFR and the star formation efficiency (SFE).   
In M1 and M2, the star formation rate peaks after roughly 1.6 $t_{\mathrm{ff}}$, and plateaus for the next free-fall time. On the other hand, in M3, the SFR continues to increase beyond $\sim$ 1.5 $t_{\mathrm{ff}}$, and is still increasing at $\sim$ 2.5 $t_{\mathrm{ff}}$. The high initial gas mass and surface density ($>$ 10$^{3}$ M$_{\odot}$pc$^{-2}$) prevent the feedback from efficiently stopping star formation~\citep[see discussion in][]{Menon2023, Polak2024}. We also verify what mass fraction of the stars is bound at 2.5 $t_{\mathrm{ff}}$, and find a bound mass fraction above 98\% for all three simulations. 
We note that the differences in SFE lead to super-linear relation between bound stellar mass and initial cloud mass, which allows us to probe a larger range of stellar masses within a cluster-forming region than suggested by the mass range of our initial clouds.

\subsection{Stellar density}

\begin{figure}[tb!]
    \centering
    \includegraphics[width=\linewidth, clip=True, trim=0cm 0cm 0cm 0cm]{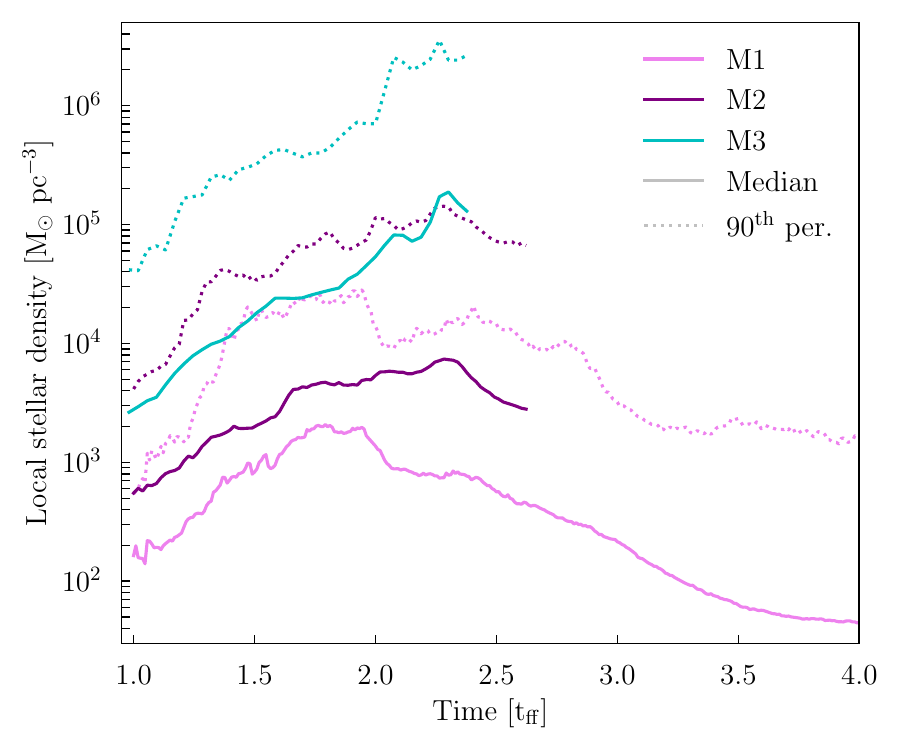}
    \caption{Median and 90th percentile local stellar density as a function of time in units of initial GMC free-fall time.}
    \label{fig:rho_tff}
\end{figure}

The differences in SFR and SFE have consequences for the density of the sub-clusters embedded within the GMC. Due to the high degree of substructure and non-sphericity of the systems, rather than calculating a global measure of density, we adopt a local measure of stellar density. We calculate the local stellar density based on the distance to stars' nearest neighbours and their masses, using
\begin{equation}\label{eq:rho}
    \rho = \frac{3}{4 \pi r_{10}^3} \sum_{i=1}^{10} m_i 
\end{equation}
where $m_i$ is the mass of the $i^{\mathrm{th}}$ nearest neighbour and $r_{10}$ is the radial distance to the 10$^{\mathrm{th}}$ nearest neighbour, where the star itself is defined as its own closest neighbour. We plot the median and 90$^{\mathrm{th}}$ percentile local stellar densities in Figure~\ref{fig:rho_tff}. 

For M1, the distribution of local densities shifts to higher values while the SFR is increasing and during its plateau. However, once star formation has slowed, the local densities start decreasing, with the median approaching 50 M$_{\odot}$ pc$^{-3}$ at late times. At all times, the stars in the highest density regions have local densities above 10$^4$ M$_{\odot}$ pc$^{-3}$ (with more than 10\% of all stars in environments with local density $> 10^4$ M$_{\odot}$ pc$^{-3}$ between 1.5 and 2 $t_{\mathrm{ff}}$), which is a density typical of local YMCs~\citep[][and references therein]{PortegiesZwart2010}. The distribution of local stellar densities even extends to 10$^5$ M$_{\odot}$ pc$^{-3}$ at early times. M2 follows similar trends but at higher densities. The decrease in densities also takes place at later times, following the longer plateau in the SFR. M3, on the other hand, exhibits very high densities: at late times, the median local density is above 10$^5$ M$_{\odot}$ pc$^{-3}$ and more than 10\% of stars have local densities above 10$^6$ M$_{\odot}$ pc$^{-3}$. This environment promotes few-body interactions and is likely to lead to binary disruption, runaway star production, and stellar mergers. 
The difference in the local density in the different cluster-forming regions demonstrates than we are probing different regimes for cluster formation with our models at different initial gas mass. Although there is only a factor of 16 difference in the initial gas mass between M1 and M3, the median local stellar density after $\sim$ 2.5 $t_{\mathrm{ff}}$ differs by more than two orders of magnitude.

\section{Evolution of binary populations}\label{sec:results}
Between any two consecutive snapshots, several effects modify the number and properties of the binaries: there is ongoing primordial binary formation (adding new systems sampled from the primordial distribution) combined with dynamical binary formation, disruption, and modification (through exchanges or encounters changing the orbital energy). In this section, we study the relative contributions of those processes to the binary population present in the cluster-forming region, as a function of time.
In Sections~\ref{subsec:time evolution fraction} and~\ref{subsec:time evolution properties}, we discuss how the binary fraction and the orbital properties evolve with time. In Section~\ref{subsec:exchanges}, we discuss how exchanges and dynamical binary formation contribute to observed changes in the distribution, while we discuss the influence of environment in Section~\ref{subsec:environment}. 

\subsection{Time evolution of binary fraction}\label{subsec:time evolution fraction}

\begin{figure}
    \centering
    \includegraphics[width=\linewidth, clip=True, trim=0.4cm 0.4cm 0cm 0cm]{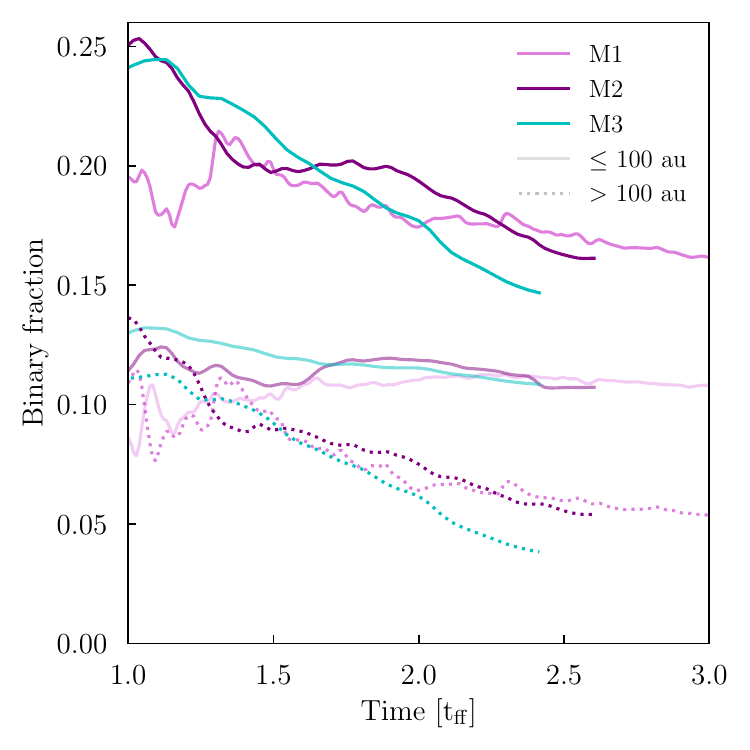}
    \caption{Binary fraction as a function of time in units of initial GMC free-fall time. Binaries with close ($\leqslant$ 100 au) and wide ($>$ 100 au) companions are shown as faint solid and dotted curves. For comparison, a fully sampled binary population with the prescription used in this work has a binary fraction of 21\%, with 12\% of the stars having a close companion and 9\% of the stars having a companion with an orbital separation above 100 au. }
    \label{fig:frac_time}
\end{figure}

We plot the binary fraction as a function of time in Figure~\ref{fig:frac_time}. 
We find that the binary fraction decreases with time in all models. Most of the decrease can be attributed to the loss of binaries wider than 100 au: all three simulations show a clear decrease in their wide binary fraction.  The most massive model, M3, is the only cluster-forming region which also shows a decrease in its close binary fraction, most likely due to the very high stellar densities it reaches.

\begin{enumerate}[(i)]
    \item For M1, which stops forming stars around 3.5 $t_{\mathrm{ff}}$, most of the decrease takes place between $\sim$ 1.5 and 2 $t_{\mathrm{ff}}$, while the star formation rate is high and the stellar density increasing. At early times, the number of stars is too small to fully sample the binary population. At late times, once star formation has stopped, the binary fraction shows no significant evolution. 
    \item In M2, the binary fraction generally decreases, but reaches a small plateau after roughly 1.5 $t_{\mathrm{ff}}$ -- this corresponds to a peak in the star formation, during which the formation of new primordial binaries balances out the dynamical disruption of wide binaries. 
    \item The decrease in the binary fraction is most obvious, and most rapid, in M3. Both M2 and M3 have binary fractions higher than the primordial binary fraction at early times (before $\sim 1.5$ $t_{\mathrm{ff}}$), due to the dynamical formation of wide binaries that are disrupted at later times.
\end{enumerate}

\subsection{Time evolution of orbital properties}\label{subsec:time evolution properties}

\begin{figure*}[!htb]
    \centering
    \begin{minipage}{.49\textwidth}
        \centering
        \includegraphics[width=\linewidth]{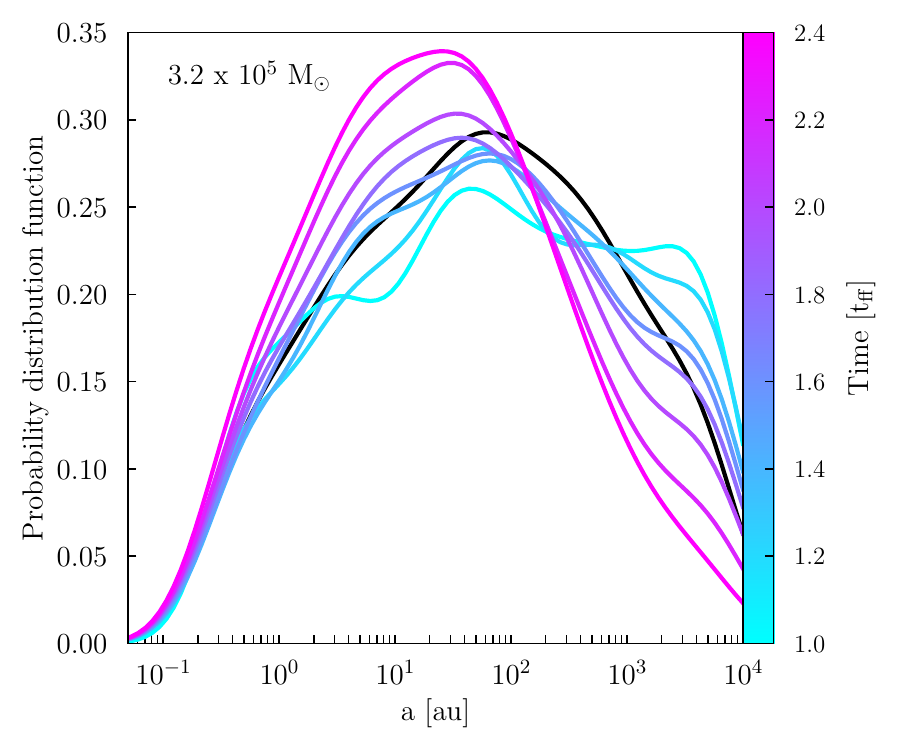}
        \includegraphics[width=\linewidth]{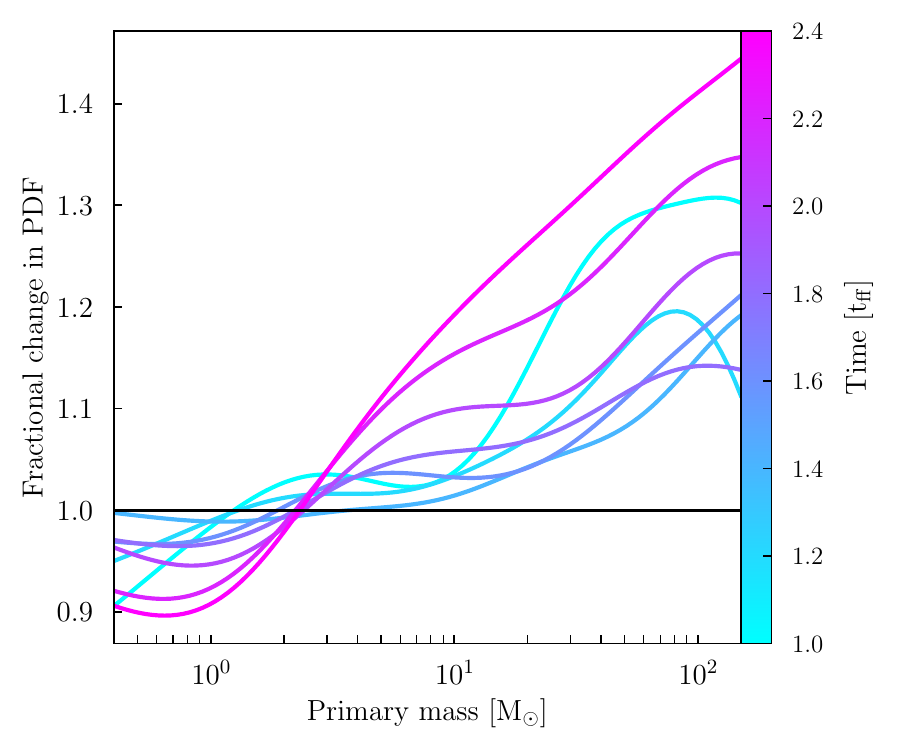}
    \end{minipage}%
    \begin{minipage}{0.49\textwidth}
        \centering
        \includegraphics[width=\linewidth]{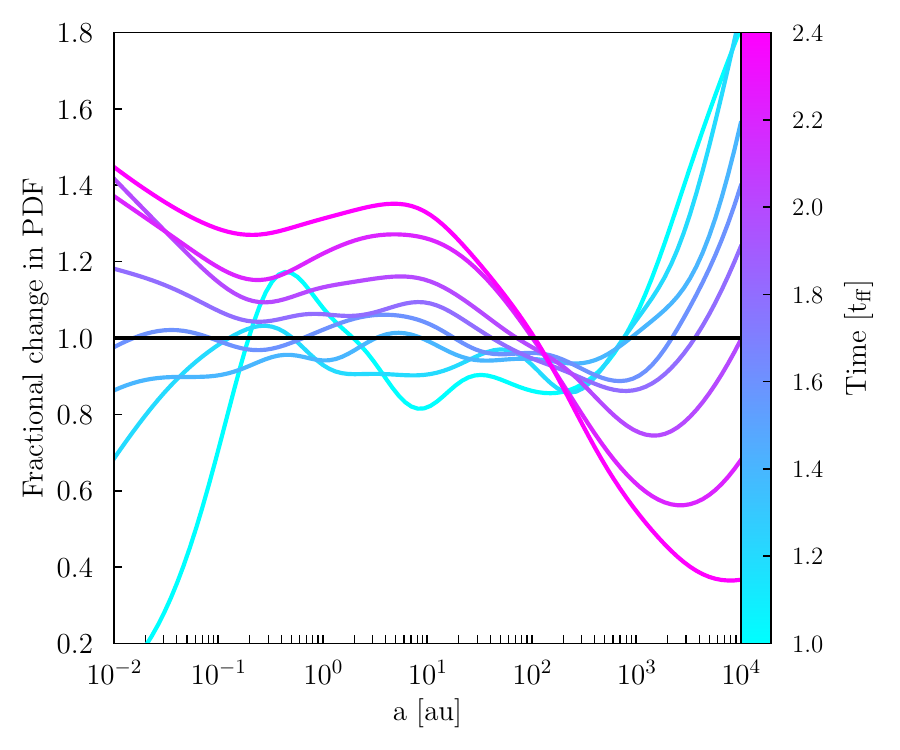}
        \includegraphics[width=\linewidth]{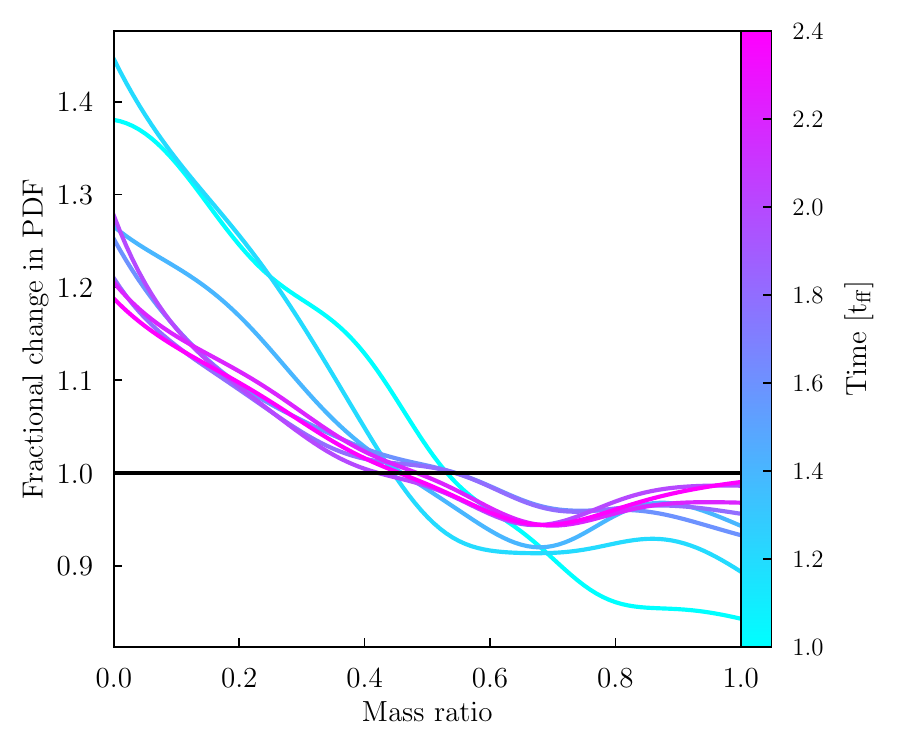}
    \end{minipage}
    \caption{\textit{Top left:} Probability distribution function of semi-major axes for M3, color-coded as a function of time in units of free-fall times of the initial cloud. The black line denotes the primordial distribution. \textit{Top right:} The same, but normalized to the primordial distribution. \textit{Bottom left:} Probability distribution function of primary masses, normalized to the primordial distribution (in black). \textit{Bottom right:} Probability distribution function of mass ratios, normalized to the primordial distribution (in black).}
    \label{fig:pdf_M3}
\end{figure*}
In this section, we use M3 as our example for the plots: it contains enough stars to fully sample the IMF and distribution of binary properties, and it exhibits the strongest signatures of the evolution of its binary population due to its high density. We do, however, note that all trends discussed below are found in all three cluster-forming regions. 

\subsubsection{Semi-major axis}
We plot the probability distribution function (PDF) of semi-major axes for M3 at different times in Figure~\ref{fig:pdf_M3} (top left), with the primordial distribution shown in black, accompanied by the distribution normalized to the primordial distribution (top right). 
The distribution of semi-major axes shifts towards smaller values with time. The largest changes in the distribution are found when the star formation rate is high and the local stellar density is increasing, between 1.5 and 2.5 $t_{\mathrm{ff}}$.
Before $\sim 1.5$ $t_{\mathrm{ff}}$, there is a peak at wide separations, associated with dynamically-formed systems. This effect is more obvious before the distribution of primordial binaries is fully sampled ($\sim 1$ $t_{\mathrm{ff}}$ for M3). Beyond $\sim 1.5$ $t_{\mathrm{ff}}$, the fraction of binary systems with separations above 100 au tends to decrease. The fraction of binaries with smaller separations increases, with a stronger increase for systems with semi-major axes below 10 au. Taken together with the results presented in Figure~\ref{fig:frac_time}, this indicates that the overall fraction of stars with at least one bound companion decreases, as well as the fraction of stars with a companion within 100 au; among stars with a bound companion, however, the fraction of bound companions within 100 au increases.

\subsubsection{Primary mass and mass ratio}
We plot the probability distribution function of primary masses and mass ratios (normalized to the primordial distribution) as a function of time for M3 in the bottom row of Figure~\ref{fig:pdf_M3} (bottom left and bottom right). The primary masses generally shift towards larger values, while the mass ratios shift towards smaller values. The fraction of binaries with OB primaries (above 2 M$_{\odot}$) increases, with the strongest fractional increase seen for the most massive O-type stars. Lower-mass binaries often have lower binding energies, and are therefore more easily disrupted. On the other hand, massive single stars tend to easily acquire bound companions~\citep[as shown in][]{Wall2019, Cournoyer-Cloutier2021}. Two effects also contribute to the shift in mass ratios. First, dynamically-formed systems, which make up about 5\% of all binaries in this simulation, tend to be paired randomly and therefore favour smaller mass ratios, due to the shape of the IMF. Low-mass binaries, which make up most of the disrupted systems (see Section~\ref{subsec:environment}), tend to have mass ratios closer to unity; their disruption therefore shifts the distribution of mass ratios towards smaller values.  

\subsection{Dynamical formation \& exchanges}\label{subsec:exchanges}

\begin{figure*}[!htb]
    \centering
    \begin{minipage}{.49\textwidth}
        \centering
        \includegraphics[width=\linewidth]{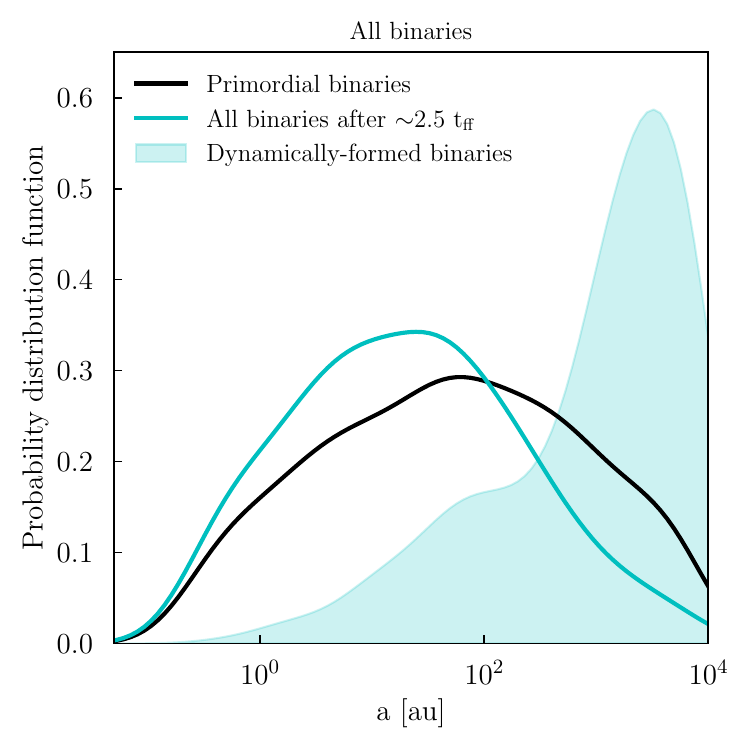}
    \end{minipage}%
    \begin{minipage}{0.49\textwidth}
        \centering
        \includegraphics[width=\linewidth]{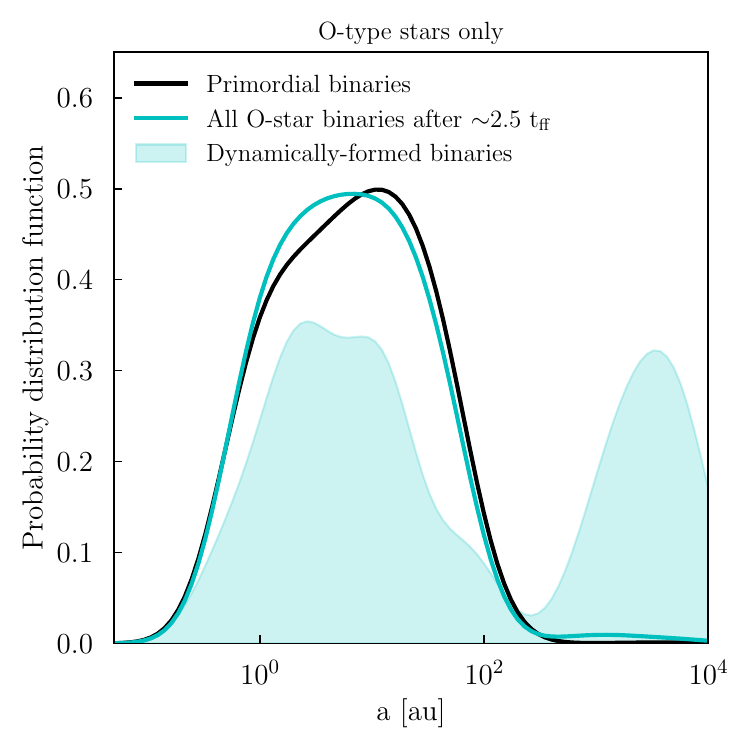}
    \end{minipage}
    \caption{\textit{Left:} Semi-major of primordial binaries, all binaries present in cluster after 2.5 $t_{\mathrm{ff}}$, and dynamically formed binaries for M3. \textit{Right:} The same, but for binaries with O-star primaries only.}
    \label{fig:a_primordial}
\end{figure*}

We can also turn our attention to the relative contributions of primordial systems -- which may be dynamically hardened or softened, or disrupted -- and systems formed dynamically, either through exchanges or by capture. In Figure~\ref{fig:a_primordial}, we compare the semi-major axis distribution of binaries present in M3 after $\sim$ 2.5 $t_{\mathrm{ff}}$ (which are primordial at $>$ 95\%) to those of the subset of dynamically-formed binaries, for all stars and for O-type stars only. For the full distribution, we find that dynamically-formed binaries tend to be much wider than primordial binaries. On the other hand, for the subset of O stars, about half of the dynamically formed binaries ($<$ 5\% of systems) have semi-major axes below 100 au; most of those are formed through exchanges. Despite their high binding energies, several primordial O-star binaries are also disrupted. Dynamical interactions during cluster formation therefore have an impact not only on the population of low-mass binaries, but also on the highest mass systems. Although O stars have a constant binary fraction (see Figure~\ref{fig:frac_stellar_mass}), several O-star binaries are modified or disrupted.

\subsection{The influence of environment} \label{subsec:environment}

\begin{figure}[!htb]
    \centering
    \includegraphics[width=\linewidth]{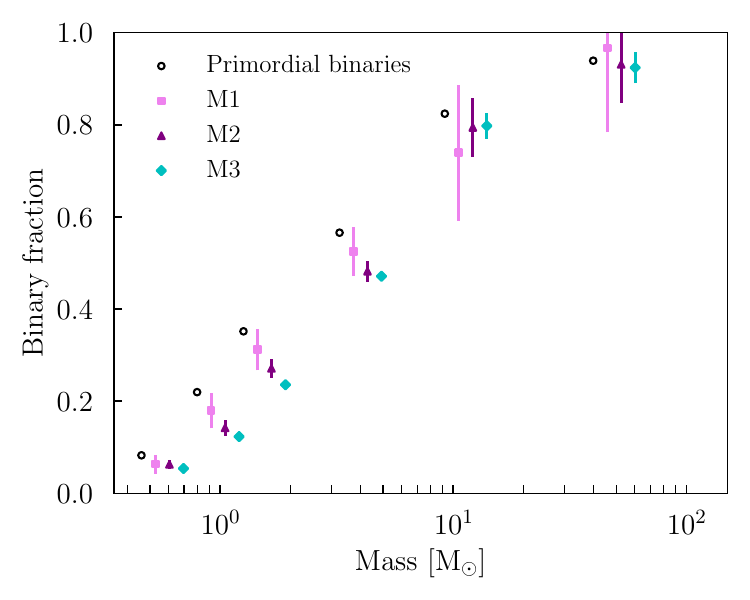}
    \caption{\textit{Left:} Binary fraction as a function of stellar mass after 2.5 $t_{\mathrm{ff}}$. The vertical lines correspond to the Poisson error, and the different runs are offset along the mass axis for readability. The mass bins used are the same ones as in Table~\ref{tab:frac_ICs}, with the center of the mass bin at the midpoint between the M1 and M2 data points.}
    \label{fig:frac_stellar_mass}
\end{figure}

We now explicitly compare the evolution of the populations of binaries in the different environments. In Figure~\ref{fig:frac_stellar_mass}, we plot the binary fraction at $\sim 2.5$ $t_{\mathrm{ff}}$ as a function of stellar mass for our simulations, compared to the primordial binaries. For M3 and M2, the binary fraction is lower than the primordial binary fraction for stellar masses $<$ 9 M$_{\odot}$, and is consistent within uncertainties with the primordial binary fraction for stellar masses $\geqslant$ 9 M$_{\odot}$. A similar trend is found for the best estimate of the binary fraction for M1, although the binary fraction is always consistent within uncertainties with the primordial distribution due to the smaller number of stars formed. This indicates clearly that the binary fraction decreases during hierarchical cluster formation for lower-mass stars, while the binary fraction for massive stars is stable. 

We also compare the probability distribution function of semi-major axes to one another and to the primordial distribution, at $\sim$ 2.5 $t_{\mathrm{ff}}$. We use the distribution of semi-major axes as evidence of the change in the binary population, as it is the metric that shows the clearest signature of change in one direction. For our statistical comparison, we use the two-sample Kolmogorov-Smirnov (KS) test. We are confident at $>$ 99\% that all three distributions have smaller semi-major axes than the primordial distribution, and that M3 has smaller semi-major axes than M2 and M1. This confirms that similar trends in the evolution of populations of binaries emerge in different cluster-forming environments, and confirms that those trends are stronger in denser, more massive cluster-forming environments. 
We find that the distributions keep diverging from the primordial distribution of semi-major axes well after the change is reliably detected. This is most obvious for M3, for which the difference keeps increasing throughout the simulation. On the other hand, the distribution of semi-major axes becomes more stable for M1 at late times, following the binary fraction. Taken together with the results from Figures~\ref{fig:SFR} and~\ref{fig:frac_time}, this confirms that most of the changes in the binary fraction and binary orbital properties take place during cluster formation, while the star formation rate is high. We present a more detailed discussion of this time evolution in Appendix~\ref{appendix:KS}

\section{Discussion}\label{sec:discussion}

\subsection{Comparison to observations}
We have simulated counterparts to cluster-forming regions with physical properties mimicking a large range of observed embedded clusters in the Local Group. Although the cluster-forming regions do not form a single, monolithic cluster within 2.5 $t_{\mathrm{ff}}$, we expect they will eventually form bound clusters of mass greater or equal to their current stellar mass.

\begin{enumerate}[(i)]
    \item M1, which is evolved until gas expulsion, forms a total stellar mass of 7.8 x 10$^{3}$ M$_{\odot}$, of which 6.9 x 10$^{3}$ M$_{\odot}$ is bound. It can therefore be treated as a simulated counterpart to M16 (8,100 stars), RCW 38 (9,900 stars) or NGC 6357~\citep[12,000 stars;][]{Kuhn2015}, which are local embedded clusters hosting massive stars.
    \item  M2, which has formed roughly 3.4 x 10$^4$ M$_{\odot}$ of stars (of which $>$ 99\% are bound), is more similar to the Arches cluster, which has a stellar mass between 2 x 10$^{4}$~\citep{Espinoza2009} and $\leqslant$ 7 x 10$^{4}$~\citep{Figer2002}, a density above 10$^{5}$ M$_{\odot}$ pc$^{-3}$ in its densest regions~\citep{Espinoza2009}, and is known to have a high binary fraction for stars more massive than 50 M$_{\odot}$~\citep{Clark2023}.
    \item  M3, with a bound mass approaching 2 x 10$^5$ M$_{\odot}$, is more massive than any YMC within the Milky Way~\citep[][and references therein] {PortegiesZwart2010}, and about twice as massive as R136 in the LMC~\citep[8.7 x 10$^{4}$ M$_{_\odot}$, ][]{Cignogni2015}, known to host several stars more massive than 100 M$_{\odot}$. 
\end{enumerate}


Our results confirm that a universal mass-dependent primordial binary fraction and distribution of orbital parameters naturally gives rise to variations in binary population properties with environment. Most of the changes in the properties of the binary population take place during the cluster assembly process, which is consistent with the decrease in binary fraction and the shift in binary properties being driven by subcluster mergers~\citep[as found in][]{Cournoyer-Cloutier2024}. This is important to take into account when comparing to observations. Several clusters for which we have resolved observations show evidence of recent or ongoing mergers between sub-clusters (e.g., Westerlund 1, \citeauthor{Zeidler2021}~\citeyear{Zeidler2021}; R136, \citeauthor{Sabbi2012}~\citeyear{Sabbi2012}, \citeauthor{Fahrion2024}~\citeyear{Fahrion2024}). However, it is hard to constrain observationally whether a cluster has undergone a recent merger, even with resolved photometry for individual stars: signatures from the shape of the cluster are erased on very short timescales~\citep{Cournoyer-Cloutier2023}, and signatures from an anisotropic distribution of runaway stars~\citep{Polak2024b} require high-quality observations away from the cluster center. The recent history of an embedded cluster therefore also likely contributes to setting its wide binary fraction and distribution of orbital properties, in addition to its density, therefore resolving the apparent inconsistency between the excess~\citep{Niu2020} or dearth~\citep{Deacon2020} of binaries observed in denser young open star clusters.

Our simulations reproduce the lower number of wide binaries for low- and solar-mass stars that is observed in dense star-forming regions~\citep[e.g.,][]{Duchene2018}, as well as the stable close binary fraction for clusters in the mass range of observed clusters in the Milky Way~\citep{Deacon2020}. For more massive clusters, however, our results suggest that the binary fraction for companions closer than 100 au also decreases during cluster assembly. Clouds with shorter free-fall times -- and therefore dense, compact clouds -- show more rapid changes to their populations of binaries during cluster formation.


\subsection{Implication for globular cluster formation}\label{subsec:GC formation}

Although it is not possible to confirm observationally whether the most massive YMCs in starbursts in the local Universe are forming through sub-cluster mergers -- let alone GCs at high redshift -- there is strong evidence from simulations that the most massive star clusters assemble from the repeated mergers of smaller sub-clusters~\citep{Howard2018, Dobbs2022, Rieder2022, Reina-Campos2024}. Observations in the local Universe and at high redshift both suggest that the most massive clusters should form from GMCs with high surface densities, and therefore short free-fall times: GMCs in local starburst galaxies have surface densities $\geqslant$ 10$^{3}$ M$_{\odot}$ pc$^{-2}$~\citep{Sun2018}, as do GMCs observed in a strongly-lensed galaxy at $z\sim$1~\citep{Dessauges-Zavadsky2023}. We therefore expect that clusters forming from those GMCs would exhibit similar behaviour to that of our M3 model, in which the distribution of binaries is strongly modified at early times in the cluster formation process.



GCs tend to have low binary fractions~\citep[$\lesssim$ 10\% assuming a field-like mass ratio distribution,][]{Milone2012}, for stars with masses $\lesssim$ 0.8 M$_{\odot}$. This corresponds to our lowest mass bin in Figure~\ref{fig:frac_stellar_mass}, and shows good agreement with our calculated fraction of systems with semi-major axes smaller than 10,000 au. 
If the primordial binary population is set by the physics of core and disk fragmentation -- and therefore the same for cluster-forming environments of the same metallicity -- then the hierarchical formation of massive clusters could be sufficient to explain the low binary fraction observed for low-mass stars in old GCs, while allowing for the high binary fraction for massive stars in YMCs. 
%


In a recent paper, \citet{Nguyen2024} showed that a population of massive binaries between 10 and 40 M$_{\odot}$, following the same distribution of orbital properties as our primordial distribution, loses about 25\% of its initial mass as pre-supernova ejecta. This cool ejecta has the right abundances~\citep[as originally shown by][]{deMink2009} to explain the light abundance variations observed in the vast majority of GCs~\citep[known as multiple populations, see][for recent reviews]{Bastian2018, Gratton2019, Milone2022}. 
In our simulations, the binary fraction for O-type stars shows very little change during the assembly process, even for the most massive, densest cloud. The semi-major axes however tend to shift to smaller values, and exchanges are ubiquitous, including for originally tight systems. This suggests that~\citet{Nguyen2024} may have underestimated the number of short-period systems present in a massive cluster, and therefore the amount of cool, enriched ejecta. We also suggest that more massive clusters should host even more close binaries than M3, in agreement with the observations that show a stronger signal of enrichment in more massive clusters. As the close binary fraction of massive stars is insensitive to metallicity~\citep{Moe2019}, the primordial binary properties for close, massive binaries should be similar for GCs forming at low metallicity and for massive stars observed in the local Universe. Our simulations thus indicate that GCs very likely hosted rich populations of close, massive binaries during their formation, supporting massive interacting binaries as a possible source of enriched material for multiple populations.


\section{Conclusions}\label{sec:conclusions}
We have conducted simulations of young massive cluster formation within GMCs with masses 2~x~10$^4$, 8~x~10$^4$ and 3.2~x~10$^5$ M$_{\odot}$, and studied how populations of binary stars evolve during the cluster formation process. We have found that the binary fraction and the distribution of the binaries' orbital properties changes faster and more strongly in more massive and denser clouds. This tendency is exacerbated by the nonlinear relationship
between initial gas mass and final stellar density that we find. We summarize the key results below. 

\begin{itemize}
    \item The binary fraction decreases rapidly in all our simulations while the star formation rate and the local stellar density are increasing. When the star formation rate and the local stellar density decrease, the binary fraction stabilizes. 
    \item A similar trend is found for the changes in the distributions of orbital properties, due to a combination of binary disruption, exchanges, and dynamical binary formation, along with ongoing star formation. The clearest trends with time are seen for the semi-major axis, which shifts towards smaller values throughout the cluster assembly process.
    \item The decrease in the binary fraction is driven by a decrease in the wide ($>$ 100 au) binary fraction, although the most massive, densest cluster-forming region also shows a decrease in its close binary fraction.
    \item For the most massive, densest cluster-forming region, the distribution of semi-major axes becomes measurably different from the primordial distribution after about 1.5 $t_{\mathrm{ff}}$, despite the ongoing rapid star formation after this point. On the other hand, for lower density environments, the distribution takes a longer time to become measurably different from the primordial distribution, due to the less concentrated star formation. 
    \item The binary fraction does not change for O-type stars, and the distribution of O stars only shows a small shift towards smaller semi-major axes. Individual systems, including ones with very tight orbits, can however be modified, for example via exchanges with other systems. 
\end{itemize}

We have found that populations of binaries evolve during clustered star formation within GMCs, and that a universal field-like primordial distribution can naturally explain the observed trends with cluster mass and density for Galactic clusters. Changes are more rapid and stronger in more massive, denser GMCs, which can naturally explain the differences in binary fractions and binary orbital properties observed in different clustered environments. While the present paper investigates the evolution of the population of binaries during star cluster formation, binaries may also impact their host cluster. In future papers in this series, we will investigate runaway stars and stellar mergers in clusters with realistic populations of binaries, as well as the effects of binary stellar evolution on star formation within cluster-forming regions (Cournoyer-Cloutier et al. in prep.).

\begin{acknowledgments}
CCC is supported by a Canada Graduate Scholarship -- Doctoral (CGS D) from the Natural Sciences and Engineering Research Council of Canada (NSERC). AS and WEH are supported by NSERC. This research was enabled in part by support provided by Compute Ontario (\url{https://www.computeontario.ca/}) and the Digital Research Alliance of Canada (\url{alliancecan.ca}) via the research allocation FT \#2665: The Formation of Star Clusters in a Galactic Context. Some of the code development that facilitated this study was done on Snellius through the Dutch National Supercomputing Center SURF grant 15220.
M-MML, BP, and EA were partly supported by NSF grant AST23-07950. EA was partly supported by NASA grant 80NSSC24K0935. SMA is supported by an NSF Astronomy and Astrophysics Postdoctoral Fellowship, which is supported by the National Science Foundation under Award No. AST24-01740; SMA also received support from Award No. AST20-09679.
\end{acknowledgments}

%






\appendix

\section{Statistical test of semi-major axis evolution}\label{appendix:KS}

\begin{figure}
    \centering
    \begin{minipage}{.49\textwidth}
        \centering 
        \includegraphics[width=\linewidth, clip=True, trim=0cm 0cm 0cm 0cm]{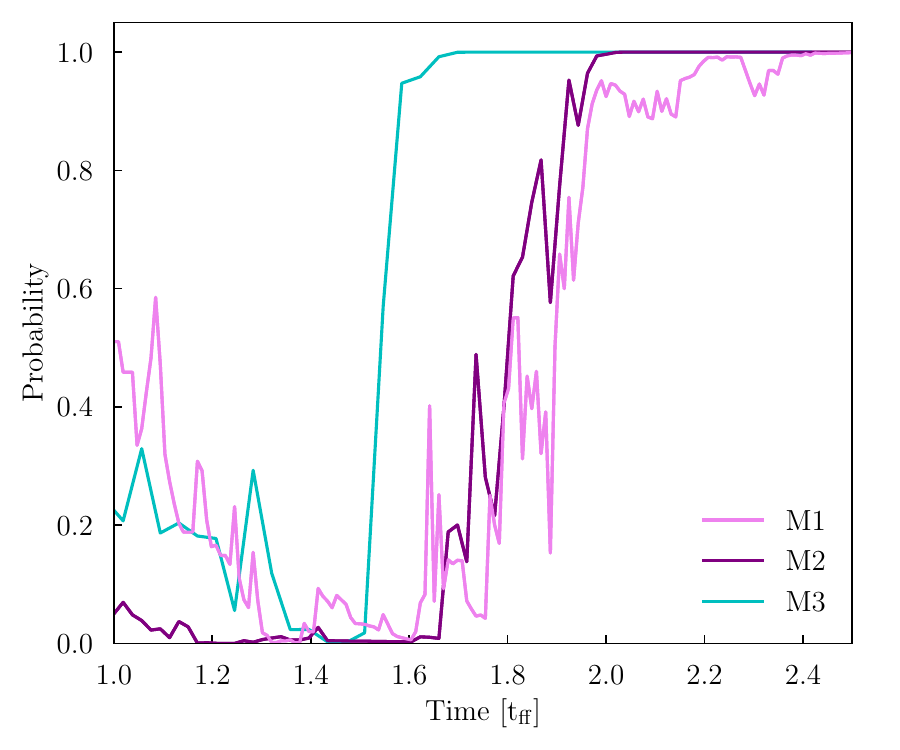}
    \end{minipage}%
    \begin{minipage}{0.49\textwidth}
        \centering
        \includegraphics[width=\linewidth, clip=True, trim=0cm 0cm 0cm 0cm]{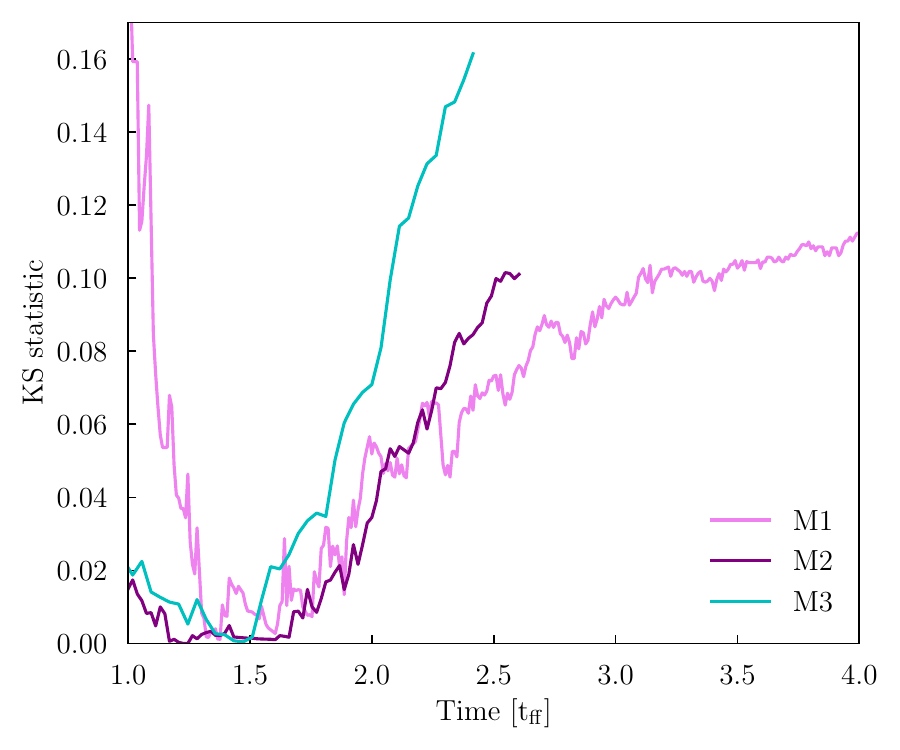}
    \end{minipage}
    \centering
    \caption{\textit{Left:} Probability that the semi-major axes are smaller than in the primordial distribution, as a function of time. M1, M2, and M3 consistently show a difference relative to the primordial distribution after 2.37, 1.98 and 1.66 $t_{\mathrm{ff}}$ (2.51, 1.05 and 0.44 Myr). \textit{Right:} KS statistic measuring the amount of change from the primordial distribution, as a function of time.}
    \label{fig:pvalue}
\end{figure}

We compare the distribution of semi-major axes in all of our simulations, at every snapshot, to the primordial distribution. We do so to investigate at what time changes to the population of binaries can be reliably measured in different environments, and whether the population continues to evolve beyond this point.
In Figure~\ref{fig:pvalue}, we plot the probability, as a function of time, that the distribution of semi-major axes has shifted towards smaller values compared to the sampled distribution. The probability is calculated from the KS test, as 1-$p$. 
The probability obtained from the KS test, however, only answers the question of whether the semi-major axes are smaller than in the primordial distribution, but does not measure by how much the distributions differ. The KS statistic itself, however, provides a measure of the difference between the two distribution. We also plot it as a function of time in Figure~\ref{fig:pvalue}. In all cases, we compare the distribution of binaries present within a cluster-forming region to the primordial distribution. We emphasize that the primordial distribution is not an initial distribution, but rather the distribution for newly-formed binaries; at any given time, the observed binary population arises from the combined contributions of primordial binary formation and the effects of dynamics.

At early times, both the probability and the KS statistic are non-zero in all simulations, due to the small number of stars formed. They approach zero around $\sim 1.5$ t$_{\mathrm{ff}}$, when the combination of stellar dynamics and new star formation result in a binary population that is very similar to the primordial population. This effect is strongest for M1, which has the fewest stars. At later times, the effects of stellar dynamics start to dominate over the formation of primordial binaries, and changes in the binary population become detectable.

For all three clusters, the probability stabilizes at $>$ 99\%. It reaches this value earlier for M3 than for M2, and earlier for M2 than for M1. Changes in massive cluster-forming clouds are stronger, and happen more quickly than in lower-mass clouds. We also note that the lower-mass models, in particular M1, can oscillate strongly between subsequent checkpoints, due to bursts in star formation. The same effect can be seen, albeit more weakly, in M2. M3, on the other hand, goes almost directly to a probability of 100\% and remains there, despite the high star (and therefore primordial binary) formation rate. 
We can calculate a time $\tau_{\mathrm{c}}$ after which the probability is stable at $>$ 99\%, i.e., a timescale for a significant change in the binary population, for all simulations. We get values of 2.37, 1.98 and 1.66 $t_{\mathrm{ff}}$ (2.51, 1.05 and 0.44 Myr) for M1, M2, and M3.  We find that the timescale for change increases with the cloud's initial free-fall time. $\tau_{\mathrm{c}}$ increases superlinearly with free-fall time, like the stellar mass formed in the clouds: more massive, denser clouds undergo more rapid star formation and more rapid changes to their populations of binaries.

\bibliography{bibliography}{}
\bibliographystyle{aasjournal}



\end{document}